\documentclass[preprint,pre,aps,showpacs]{revtex4}
\usepackage{graphicx}
\begin{document}

\title{Phase diagram for asymmetric nuclear matter in the
multifragmentation model} 

\author{G. Chaudhuri$^1$, and S. Das Gupta$^2$}

\affiliation{$^1$Variable Energy Cyclotron Centre,
1/AF Bidhan Nagar, Kolkata700064,India} 

\affiliation{$^2$Physics Department, McGill University, 
Montr{\'e}al, Canada H3A 2T8}

\date{\today}

\begin{abstract}

We assume that, in equilibrium, nuclear matter at reduced density and
moderate finite temperature, breaks up into many fragments.  A strong
support to this assumption is provided by data accumulated from 
intermediate energy heavy ion collisions.  The break-up of hot 
and expanded nuclear matter
according to rules of equilibrium statistical mechanics is the 
multifragmentation model. The model gives a first-order phase transition.
This is studied in detail here.
Phase-equilibrium lines for different degrees of asymmetry are computed. 

\end{abstract}


\maketitle

\section{Introduction}
Nuclear matter is a hypothetical very large system of nucleons where
the Coulomb effects of protons are switched off.  Such a system is
expected to have features of liquid-gas phase transition.
We consider here the equation of state of symmetric and asymmetric
nuclear matter at temperature between 4 and 10 MeV and at less
than half of normal nuclear density.
We assume that at equilibrium at finite temperature (three to tens of MeV)
and low average density, nuclear matter breaks up into fragments, each
with normal nuclear density.  Strong support to this assumption comes
from data on heavy ion collisions but it is also supported by theoretical
modelling.  For example it can be easily shown (see section IV in 
\cite{Bhatta}), using Skyrme type interaction, that the free energy of 
uniformly stretched nuclear matter is very 
significantly lowered if the matter is allowed to split into many
fragments, each with normal nuclear density.  This is the multifragmentation
model.  We use this model to study thermodynamic properties of
nuclear matter, particularly 
phase-equilibrium lines (the lines of co-existence of liquid and
gas phases) in 
the $p-T$ plane for both symmetric and asymmetric matter.

This is an extension of the model described in our earlier work
\cite{Chaudhuri1} where 
only one kind of particles was considered.  This one kind of particles
however formed clusters whose properties were patterned after actual 
finite nuclei.  While we hope that the present article is self-contained we 
will refer to this earlier work for elucidation of some points.  There is a
large number of publications on equation of state and phase transitions
in nuclear matter.  Ref. \cite{De1} comes closest to the spirit of this
paper.  While there are quite a few common features with this work
there are also some differences and we highlight some other aspects.
Phase transition in nuclear matter using 
mean-field theory was studied over many years and we 
can not attempt an adequate bibliography here.  We mention two
papers which critically looked at asymmetric nuclear matter and
received a great deal of attention in very recent
times \cite{Muller,Ducoin}.  Both of these use mean-field theories
and overcome the difficulty of instability through Maxwell construction.
The multifragmentation approach is very different.  It is more directly
related to actual observables but in its present form it can only be trusted
in a low density regime.  But there is no need for Maxwell construction.

\section{The Formulae}

We briefly review the grand canonical model for multifragmentation
\cite{Dasgupta1}.  Let the numbers of neutrons and protons in the dissociating
system be $N_0$ and $Z_0$ respectively.  At finite temperature and in 
subnormal densities, these will break up into all possible
composites each with some neutrons($N$) and protons ($Z$)(mass number $A=N+Z$).
We always use the subscripted $N_0,Z_0$  to refer to the very large system
whose thermodynamic properties are being investigated whereas $N,Z$ refer
to composites which can be small or large.
The properties of the composites are determined by the basic two-body
interactions  These properties are utilized in the model but interactions
between composites are neglected (except through excluded volume effect;
see discussion later)
by appealing to the short range nature
of nuclear forces.  This limits the validity of the model to low densities.
Here we will restrict our investigation to densities $\rho/\rho_0$
to 0.5 or less where $\rho_0$ is the normal nuclear density.  
This is the customary practice \cite{Das1}.

If the neutron chemical potential is $\mu_n$
and the proton chemical potential is $\mu_p$, then statistical equilibrium
implies that the chemical potential of a composite
with $N$ neutrons and $Z$ protons is $\mu_nN+\mu_pZ$.  The following are the
relevant equations for us.  The average number of composites with $N$ neutrons
and $Z$ is ($\beta=1/T$)
\begin{eqnarray}
\langle n_{N,Z}\rangle=e^{\beta\mu_nN+\beta\mu_pZ}\omega_{N,Z}
\end{eqnarray}
Here $\omega_{N,Z}$ is a one body partition function for the composite 
$(N,Z)$.  It is a product of two factors;  one arising from
the translational motion of the composite and another from the
intrinsic partition function of the composite:
\begin{eqnarray}
\omega_{N,Z}=\frac{V_f}{h^3}(2\pi mT)^{3/2}A^{3/2}\times z_{N,Z}(int)
\end{eqnarray}
Here $V_f$ is the volume available for translational motion; $V_f$ will
be less than $V$, the volume to which the system has expanded at
break up (excluded volume correction). 
We use $V_f = V - V_0$ , where $V_0$ is the normal volume of  
nucleus with $Z_0$ protons and $N_0$ neutrons.  The quantity $z_{N,Z}(int)$
depends upon the intrinsic properties of the composites and contains
all the nuclear physics.

We list now the properties of the composites used in this work.  The
proton and the neutron are fundamental building blocks 
thus $z_{1,0}(int)=z_{0,1}(int)=2$ 
where 2 takes care of the spin degeneracy.  For
deuteron, triton, $^3$He and $^4$He we use $z_{i,j}(int)=(2s_{i,j}+1)\exp(-
\beta e_{i,j}(gr))$ where $e_{i,j}(gr)$ is the ground state energy
of the composite and $(2s_{i,j}+1)$ is the experimental spin degeneracy
of the ground state.  Because we are modeling a system where protons do not
carry any charges the ground state energy of $^3$He is taken to be that of the
triton and the Coulomb energy is subtracted from the experimental energy
of the alpha particle.  These modifications make insignificant changes. 
Excited states for these very low mass
nuclei are not included.  
For mass number $a=5$ and greater we use
the liquid-drop formula.  This reads 
\begin{eqnarray}
z_{i,j}(int)=\exp[-\frac{F_{i,j}}{T}]
\end{eqnarray}
Here $F_{i,j}$ is the internal free energy of species $(i,j)$:
\begin{eqnarray}
F_{i,j}=-W_0a+\sigma(T)a^{2/3}
+s\frac{(i-j)^2}{a}-\frac{T^2a}{\epsilon_0}. 
\end{eqnarray}
The expression includes the 
volume energy, the temperature dependent surface energy
and the symmetry energy.  The values of the paprameters are taken
from \cite{Bondorf1}.
The term $\frac{T^2a}{\epsilon_0}$
represents contribution from excited states
since the composites are at a non-zero temperature. For nuclei with 
$A$=5 we include
$Z$=2 and 3 and for $A$=6 we include $Z$=2,3 and 4.  For higher masses
we compute the drip lines using the liquid-drop formula above and
include all isotopes within these boundaries.

There are two equations which determine $\mu_n$ and $\mu_p$.
\begin{eqnarray}
N_0=\sum Ne^{\beta\mu_nN+\beta\mu_pZ}\omega_{N,Z}
\end{eqnarray}
\begin{eqnarray}
Z_0=\sum Ze^{\beta\mu_nN+\beta\mu_pZ}\omega_{N,Z}
\end{eqnarray}
We want to point out the following feature of the grand canonical model.  
In all $\omega_{N,Z}$'s in the sum in the above two equations,
there is one common value  for $V_f$ (see eq.(2)).  We really solve 
for $N_0/V_f$
and $Z_0/V_f$.  The values of $\mu_n$ or $\mu_p$ will not change if we, say,
double $N_0, Z_0$ and $V_f$ simultaneously provided the number of
terms in the sum is unaltered.  We then might as well say that when
we are solving the grand canonical equation we are really solving for
an infinite system (because we know that fluctuations will become
unimportant) but this infinite system can break up into only certain
kinds of species as are included in the above two equations.  Which
composites are included in the sum is an important physical ingredient
in the model but intensive quantities like $\beta,\mu$ depend not
on $N_0, Z_0$ but on $N_0/V_f$ and $Z_0/V_f$.  

The choice of which nuclei are included in the sum of the right hand side
of eqs. (5) and (6) needs further
elucidation.  We can look upon the sum on $N$ and $Z$ as a sum
over $A$ and a sum over $Z$.  In principle $A$ goes from 1 to $\infty$ and
for a given $A$, $Z$ can go from 0 to $A$.  Here for a given $A$
we restrict $Z$ by the drip lines.  Comparisons with calculations
where restrictions by drip lines are not imposed (as in the
Copenhagen statistical multifragmentation model) showed that
restrictions by driplines generate imperceptible differences
\cite{Botvina1}.  De et al \cite{De1} reach similar conclusion.
Let us now consider the restriction on $A$.  In principle this should be
$\infty$ but for practical calculations one needs to restrict this
to a maximum value that we label as $A_{max}$.  Earlier calculations
with one kind of particles showed that with $A_{max}=200$ features
of liquid-gas phase transition are not revealed (see Fig.14 in
\cite{Das1}) but a high value of $A_{max}$ at 2000 produces a nearly
a perfect model of phase transition (elaborated in much
larger detail in \cite{Chaudhuri1} and in \cite{De1}).  

\section{signatures of phase transition in the model}
We now demonstrate  that the multifragmentation model predicts first order
phase transition.
There are three signatures we will dwell on.  Pressure in the model
is given by $p=T\frac{\sum n_{N,Z}}{V_f}=T\frac{\sum n_A/A_0}{V_f/A_0}=
T\rho_f\frac{\sum n_A}{A_0}$.  We plot results as function of 
$\rho$ rather than $\rho_f$ the connection being $\rho_f=\frac{\rho\rho_0}
{\rho_0-\rho}$.  We have $\rho=\rho_n+\rho_p$.  We need an asymmetry 
parameter.  We use both $N_0/Z_0$ and $\omega=\frac{N_0-Z_0}{N_0+Z_0}$.

We show in Fig.1 $p-\rho$ curves for $N_0/Z_0=1.4$ where the values of
$A_{max}$ are 200, 400, 600, 800 and 1000.  
The temperature used is $T=6.5$ MeV.
For all five choices of $A_{max}$ pressure against $\rho$ initially rises
quite sharply and then flattens out considerably.  The initial
stage of fast rise of pressure with density is the gas phase.  Here the 
results do not matter whether $A_{max}$ is 200, 400 or larger.  The reason 
will become clearer later (it is explained in detail in \cite{Chaudhuri1}).
The flattening which follows depend on $A_{max}$ but above a large
enough value of $A_{max}$ will not change.  For one kind of particles
this is reached around 2000 \cite{Chaudhuri1}.  However, the choice
of $A_{max}$=600 is good enough for at least a semi-quantitative estimate
of various thermodynamic properties of nuclear matter and we will
present results for this value although we did some calculations with
other choices of $A_{max}$ also.  The flattening happens slightly
beyond $\rho/\rho_0$=0.1.  We show results up to
$\rho/\rho_0=0.5$ arguing that the excluded volume correction for 
interactions between composites becomes worse with increasing density.

The rise of pressure at small density followed by a flattening of $p$
with increasing density is a signature of first-order liquid-gas transition. 
We have shown results for T=6.5 MeV.  Beyond a certain temperature
the flatness will disappear showing that there is no
more phase transition in the domain $\rho/\rho_0\le 0.5$.  Similarly
the flattening of $p$ disappears beyond some value of $N_0/Z_0$.
The liquid-drop parameters we are using give us for large nuclei
the drip-line at $N/Z$ (and of course $Z/N$) about 2.  Hence for
larger values of $N_0/Z_0$ the system can not stay together even at
$T$=0.  Then we will have a system which has a bound core
but always many free nucleons which will dominate the thermodynamic
properties of the system.  This is not a system we want to study.
Hence in this work we constrained ourselves to system whose $N_0/Z_0$
spans 1.0 to 1.8.  The upper limit is indeed a highly asymmetric system.

Below the density where phase transition sets in, the system is
in pure gas phase. At phase transition point some liquid will be
formed and the fraction of nucleons in the liquid phase will grow at the
expense of the gas particles as the density increases.  
This can actually be followed.  One also
gets a functional definition of what constitutes the gas particles.
Here our identification is very different from what is concluded in
\cite{De1} but very similar to what is found in our earlier work
with one kind of particles \cite{Chaudhuri1}.
 
Lastly, in one component model there is just one $\mu$ which stays constant
throughout the co-existence region.  Now there are two chemical potentials
$\mu_n$ and $\mu_p$.  How do they behave?

\section{What constitutes the gas and what constitutes the liquid?}

The quantity $<n_A>\equiv \sum_{N+Z=A}<n_{N,Z}>$ is the average number
of composites with mass number $A$.  The quantity $ A<n_A>/A_0$
gives the fraction of particles tied up in composites with mass number
$A$.  This is plotted in Figs.2 and 3 for $N_0/Z_0$=1.0 and $N_0/Z_0$=1.8
respectively.  First concentrating on Fig.2 (T=6.5 MeV) we see that at 
density $\rho/\rho_0$=0.1 the nucleons are bound in composites $\le$50.
These particles constitute the gas phase.  At density $\rho/\rho_0$
=0.3 some heavy composites with $A\approx A_{max}$ begin to form
and the probability of such heavy particles (with $A$ between $A_{max}$
and $A_{max}$-100) begins to increase (at the expense of the light particles) 
as the density increases.  This is a clear evidence of co-existence.
We thus consider light particles ($A\le 70$) to be gas and heavier
particles (with $A$ between $A_{max}$ and $A_{max}$-135) to be liquids.
Fig.3 displays similar physics but for $N_0/Z_0$=1.8 : all gas particles
at $\rho/\rho_0$=0.1 and mixture of gas and liquid at $\rho/\rho_0$=0.38.   

We note that even the gas phase in fragmentation model is quite complicated.
It is not just neutrons and protons but other light nuclei as well.
In addition, during co-existence the isotopic content of the gas phase changes
continuously as the volume of the container, i.e. density 
$\rho/\rho_0$ changes.  This is called isospin fractionation and is well-known
in literature.  We will briefly come back to this aspect later.

\section{Chemical Potentials}
In numerical work involving one kind of particles only 
\cite{Chaudhuri1} it was demonstrated that
in the limit of $A_{max}\rightarrow \infty$ a constant value of
$\mu$ will be achieved in the co-existence region.  This value
could be obtained by extrapolation.  In the present case
there are two chemical potentials.
For $N_0/Z_0\ne 1$, $\mu_n\ne\mu_p$.  For $N/Z$=1.4 
and temperature 6.5 MeV we show in Fig.4 the evolution of $\mu_n$
and $\mu_p$ as a function of density.  One notices that both $\mu_n$ 
and $\mu_p$ change rapidly in the gas phase and then tend to a constant value.
In the limit $A_{max}\rightarrow \infty$ we expect they will become constants.
We also plot in the same figure
$\mu\equiv\frac{N_0}{N_0+Z_0}\mu_n+\frac{Z_0}{N_0+Z_0}\mu_p$.  The $\mu$ so
defined has a meaning at the three limits:-1, 0 and +1 for asymmetry
parameter $\omega=\frac{N_0-Z_0}{N_0+Z_0}$ and it is interesting to note
that $\mu$ tends to constant value faster than either $\mu_n$ or $\mu_p$. 

\section{Co-existence Lines}
Figure 5 shows that as the temperature increases phase co-existence
finally disappears (from the region $\rho/\rho_0\le 0.5$).  We have
shown this for $N_0/Z_0$=1.0 but this is also true for asymmetric
systems provided the asymmetry is not too large as explained earlier.

We show in Fig.6 the $p-\rho$ curves at $T$=6.5 MeV for three systems
with $N_0/Z_0$=1, 1.4 and 1.8.  We can identify from the figure points
$A, B$ and $C$ on these curves where co-existence sets in.  The values
of pressure at these points give us $p$ values for co-existence at this
temperature for these $N_0/Z_0$ values..  This is not strictly correct.  
The values of $p$ 
increase slightly as one moves towards higher density.  This is because
with $A_{max}$=600 we have not reached asymptotic limits yet.  However,
this is adequate for our purpose.  Repeating this analysis for different
temperatures we get co-existence lines in $T-p$ plane for nuclear matter
with different asymmetries (Fig.7).  Notice that while the co-existence
lines for differing asymmetries are different they are quite similar.
As usual, points to the left and above the co-existence lines are
in the liquid phase and points to the right and below are in the gas phase.

The highest point of a co-existence line in the $T-p$ plane usually
identifies critical values $T_c,p_c$ \cite{Reif}.  This is not true
in Fig.7.  As we consider higher temperatures, points A, B and C
(Fig.6) will move to the right and up.  They will reach the $\rho/\rho_0=
0.5$ line.  These define the end-points $T,p$ in Fig.7.
We do not continue to higher densities as
the simple approximation  of excluded volume as a means of incorporating
interactions between clusters becomes progressively worse.  If we accept the
validity of the simple multifragmentation model in the region $\rho/\rho_0
\le 0.5$ we will have to conclude that the critical point does not exist in 
the region $\rho/\rho_0\leq 0.5$.  The same conclusion can be guessed from
other published work.  Multifragmentation with one kind of particles 
was also studied by Bugaev et al \cite{Bugaev}.  This is the same 
physics problem
as considered in \cite{Chaudhuri1} but treated in a different mathematical
framework and these authors considered all densities, not just 
$\rho/\rho_0\le 0.5$.  They found that one can identify a critical point
at $T=T_c=18.0$ MeV, $\rho/\rho_0=1$ and $p_c=\infty$.  At very high 
pressure the model must break down but this is an additional confirmation
that the simple multifragmentation model in the domain $\rho/\rho_0\le 0.5$
does not contain the critical point.

\section{Isothermals in a two-component system}

Figure 6 gives the isothermals for $N_0/Z_0$=1, 1.4 and 1.8 at 6.5 MeV
temperature.  Drawing isothermals for fixed $N_0/Z_0$ is physically
relevant.  We are assuming that we have a very large system with 
given numbers $N_0$,$Z_0$ whose volume can change depending upon the
physical conditions it is subjected to.  If we want to study a different
asymmetry we change $N_0/Z_0$ accordingly and repeat the calculation.
To have a complete knowledge, calculations should be done for all
relevant $N_0/Z_0$.  The most asymmetric system we study is $N_0/Z_0$=1.8.
Of course, since we have no Coulomb force the system with $N_0/Z_0=\alpha$
has the same thermodynamic properties as the system with $Z_0/N_0=\alpha$.

It is however instructive to consider isothermals of two-component
systems in a more general fashion.  In a one-dimensional system there
is only one density and an isotherm is a line in the $p-\rho$ plane. 
Now we have two densities $\rho_n$ and $\rho_p$ and isothermals
become surfaces.  Let $\rho_p$ be the x-axis, $\rho_n$ the y-axis
and $p$ the z-axis, the equation of state at a given temperature is
a surface in this space.  A projection of this surface in two
dimensions can be made but for a quantitative study it is more
convenient to present contours of constant $p$ in $\rho_p,\rho_n$
plane.  Such a plot is shown in Fig.8.  We consider pressure contours
in the region bounded by $\rho_p=1.8\rho_n, \rho_n=1.8\rho_p$ and
$(\rho_n+\rho_p)/\rho_0\leq 0.5$.  The reasons for choosing these
boundaries were explained before.

Roughly speaking, the contours are either largely radial or circular.
Let us first consider an uninteresting gas.  We assume it consists of 
only neutrons and protons and unlike in the present problem does not
form composites.  In such a case constant pressure curves would be
$\rho_n+\rho_p$=constant and these would be straight lines making angle of
$\pi/4$ with the x and y axes.  Instead we see at low $\rho_n$ and $\rho_p$
(when one has a gas phase only)
not straight lines but more like concentric circles.  
This is because pressure is directly
proportional to multiplicity (section III) and multiplicity is a function
of asymmetry.  In our case, composites are present
in the gas phase and the number of composites depend upon the
asymmetry of the system.  This causes constant pressure contours 
in the gas phase to
bend from straight lines.  We skip the details why the lines become
like circles. We now try to explain other pressure contours which are
largely radial.  For this, refer back to Fig.6.  We have mentioned before 
that in the limit $A_{max}\rightarrow\infty$ the $p-\rho$ curves would
have zero slopes to the right of
points A, B and C on the isothermals.  In such a case constant pressure 
contour would move exactly radially inwards from the boundary 
$\rho/\rho_0=0.5$,
would later leave the radial pattern, bend and finish at the boundary
$\rho_n=1.8\rho_p$ or $\rho_p=1.8\rho_n$ whichever is appropriate.  Similar
behaviour is seen in Fig.8.  Thus radial pressure contours 
reflect regions of co-existence.

As another example we show in Fig.9 the pressure contours at $T$=7.5
MeV.  Except near the edges of the boundaries, pure gas phase is seen.

\section{Isospin Fractionation}
We illustrate isospin fractionation in multifragmentation model through
an example.  Consider multifragmentation of a neutron rich system:
$N_0/Z_0$=1.4 and temperature $T$=6.5 MeV.  At low density the system
is in pure gas phase.  Following section IV, the gas phase consists of light
particles with $A\leq 70$ and the liquid phase consists of particles with
$A$ between $A_{max}$-70 and $A_{max}$.  At higher density, both gas and 
liquid phases
are seen (Figs. 2 and 3).  In the present example
with $N_0/Z_0$=1.4,  
we expect that during co-existence
the neutron to proton ratio in the gas phase will rise above 1.4 and 
the neutron to proton ratio in the liquid phase will 
fall below 1.4.  The reason for this is the symmetry energy which 
preferentially favours
formation of larger clusters closest to maximum stability (i.e.,$N=Z$).
This rise of neutron to proton ratio in the gas phase is illustrated in Fig.10.
Co-existence sets in a little beyond
$\rho/\rho_0$=0.1.  Till that point is reached the neutron to proton ratio
in the gas phase is at 1.4, the ratio of the parent system.  Then as the
density increases the ratio increases.

Fig.10 also shows that even at very low density the ratio of unbound
single neutrons to unbound single protons rises very rapidly.  But this has
got nothing to do with what is called isospin fractionation.  In fact
nothing special happens to this ratio when co-existence sets in.  It is
only if the gas phase is considered to be not just single nucleons but includes
light particles as well that isospin fractionation becomes an order parameter
if $N_0/Z_0 \neq$ 1.

In the present example, at $\rho/\rho_0=0.35$ the neutron to proton ratio 
in the gas phase is 1.485.  In the liquid phase it is 1.375.

Isospin fractionation in mean-field theories is treated in
\cite{Muller,Ducoin}.  Calculations in the lattice gas model
can be found in \cite{Chomaz1}.  

\section{Summary}
The multifragmentation model, so useful for fitting experimental data
in intermediate energy collisions, leads naturally to a model
of phase transition for nuclear matter.  In a range of temperature and
density first order phase transition occurs. The gas phase and 
the liquid phase can be clearly identified. This is really remarkable.
The model of nuclear multifragmentation may be unique in this respect.
The gas phase consists of 
light nuclei with $A$ up to about 70.  Besides these gas  particles,
there are large blobs
of matter (liquid) with mass numbers close to $A_{max}$ with 
$A_{max}\rightarrow\infty$.  
The model is appropriate at subnormal
nuclear density.  Modifications of the simple model are needed
to extend the model to higher density but this may not be easy.
 
Actual nuclear systems as created in heavy ion collisions are
finite and in addition have Coulomb forces.  This makes identification
of signals which are finger prints of phase transition difficult.
This continues to be the subject of intense study 
and there is large volume of literature but this is outside the scope
of the present article.

\section{Acknowledgement}
This work is partially supported by the Natural Sciences and Engineering
Research Council of Canada.  Work reported in section VII had its 
origin from discussions with F. Gulminelli during a past collaboration.
We acknowledge a communication with S. Samaddar.  We acknowledge generous 
help from J. Gallego for computation.

\pagebreak

\begin{figure}
\includegraphics[width=5.0in,height=5.5in,clip]{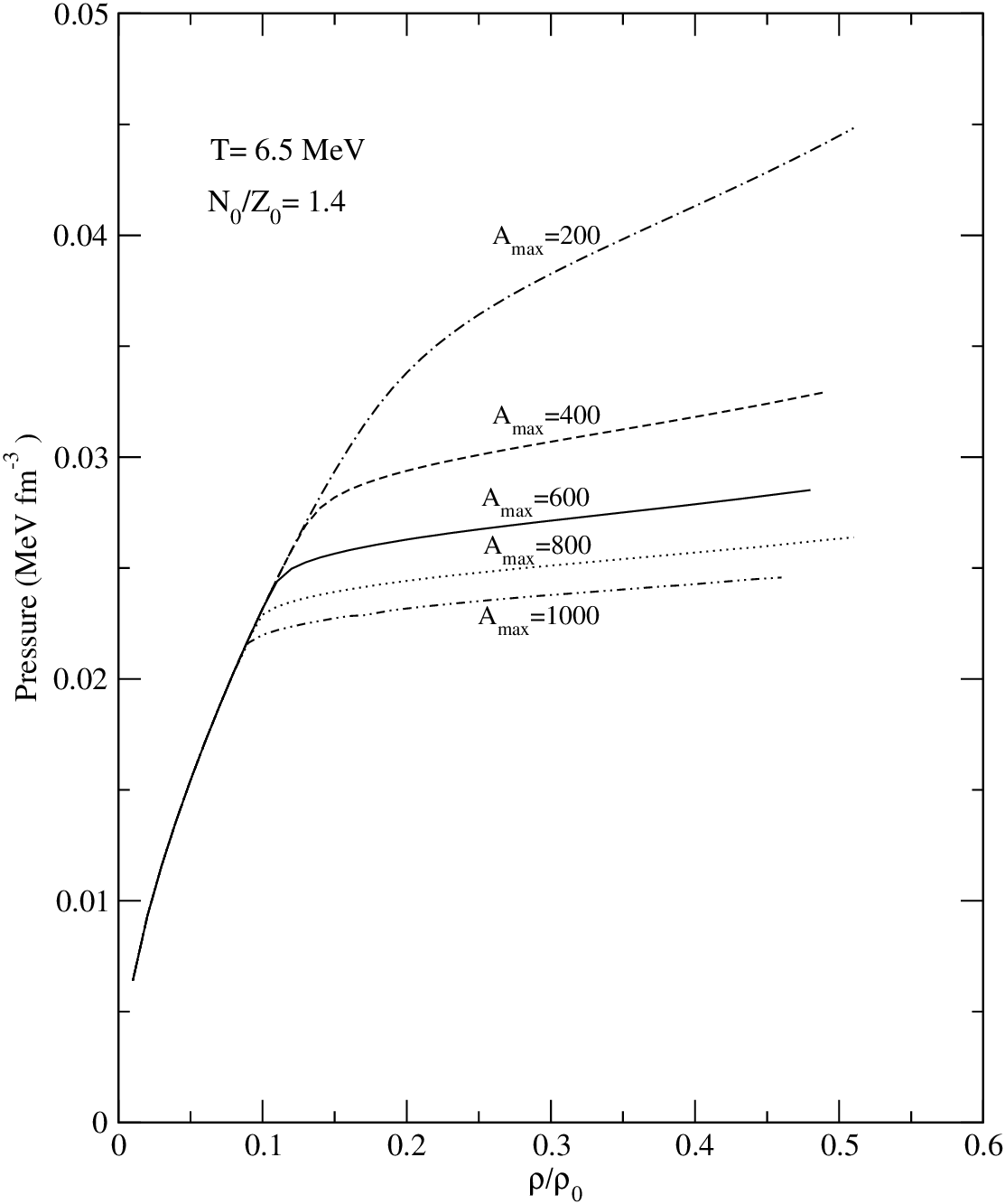}
\caption{Pressure-density curves for $N_0/Z_0=1.4$ and $ T=6.5$ MeV, where the 
values of $A_{max}$  used are 200, 400, 600, 800 and 1000.  
Note that in the region
of fast rise of pressure with density results are insensitive to the value
$A_{max}$.  In the high density side pressure appears to approach a constant
value as a function of density as the the value of $A_{max}$ is increased.
}
\label{Fig. 1}
\end{figure}

\begin{figure}
\includegraphics[width=5.0in,height=5.5in,clip]{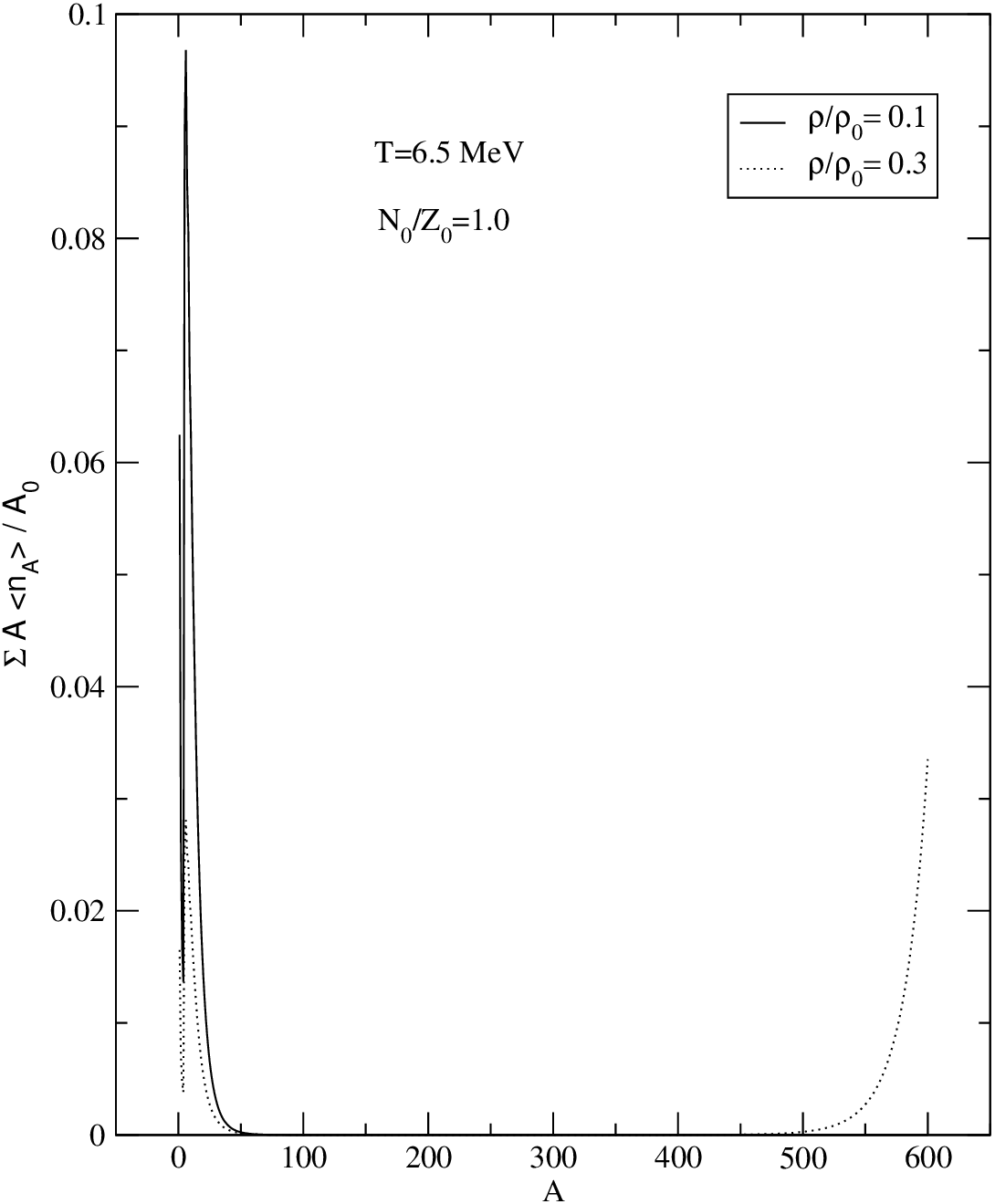}
\caption{Plot of $A<n_A>/A_0$ as a function of the mass number A for
$N_0/Z_0=1.0$ and $ T$=6.5 MeV. The solid line gives the distribution of
composites at $\rho/\rho_0=0.1$. There are practically no heavy particles, 
none above $A$=70.  This is pure gas phase.  The  dotted line is at 
$\rho/\rho_0=0.3$.  Now there are both light and heavy ($A\geq 500$)
particles.  This is co-existence.  Here and in the rest of the figures
we used $A_{max}$=600.}
\label{Fig. 2}
\end{figure}

\begin{figure}
\includegraphics[width=5.0in,height=5.5in,clip]{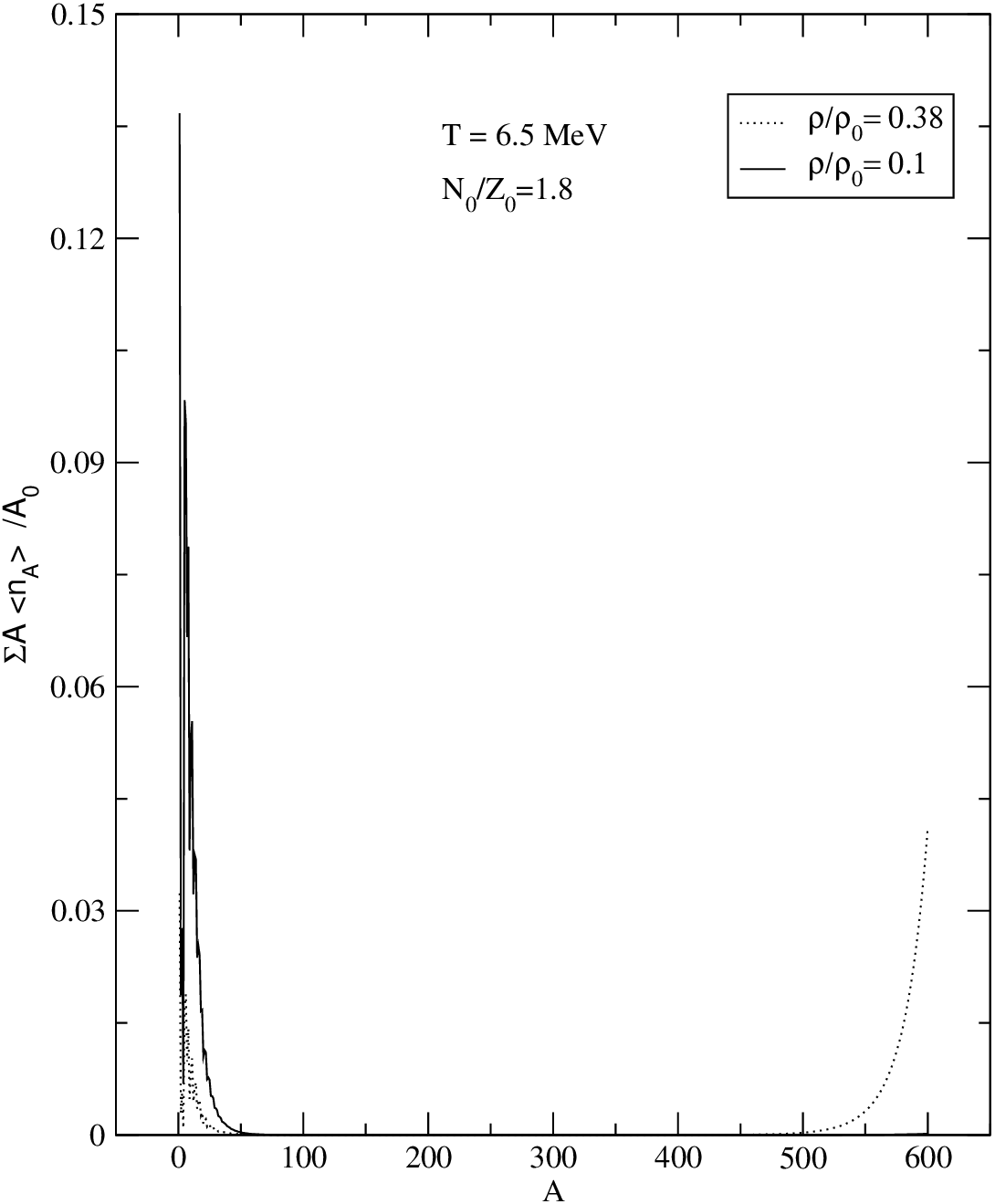}
\caption{Same as Fig.2 except that it is for $N_0/Z_0=1.8$ and the dotted line is
for $\rho/\rho_0=0.38$.}
\label{Fig. 3}
\end{figure}

\begin{figure}
\includegraphics[width=5.0in,height=5.5in,clip]{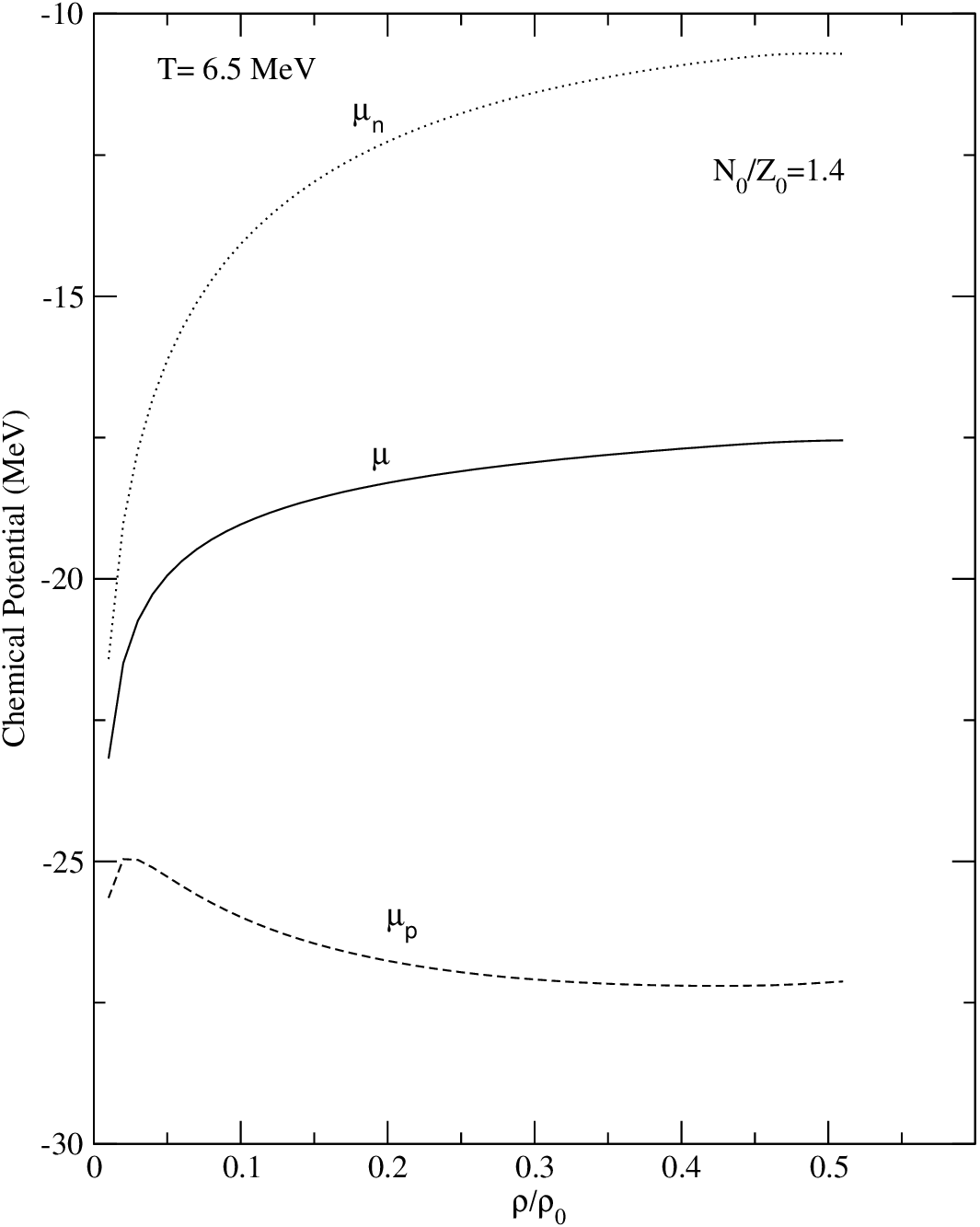}
\caption{Plot of chemical potential as function of density for $N_0/Z_0=1.4$ and $
T=6.5$ MeV. The dotted line is the  neutron chemical potential $\mu_n$, 
the dashed
line is the proton chemical potential $\mu_p$ and the solid line is 
$\mu=\frac{N_0}{A_0}\mu_n+\frac{Z_0}{A_0}\mu_p$. 
}
\label{Fig. 4}
\end{figure}

\begin{figure}
\includegraphics[width=5.0in,height=5.5in,clip]{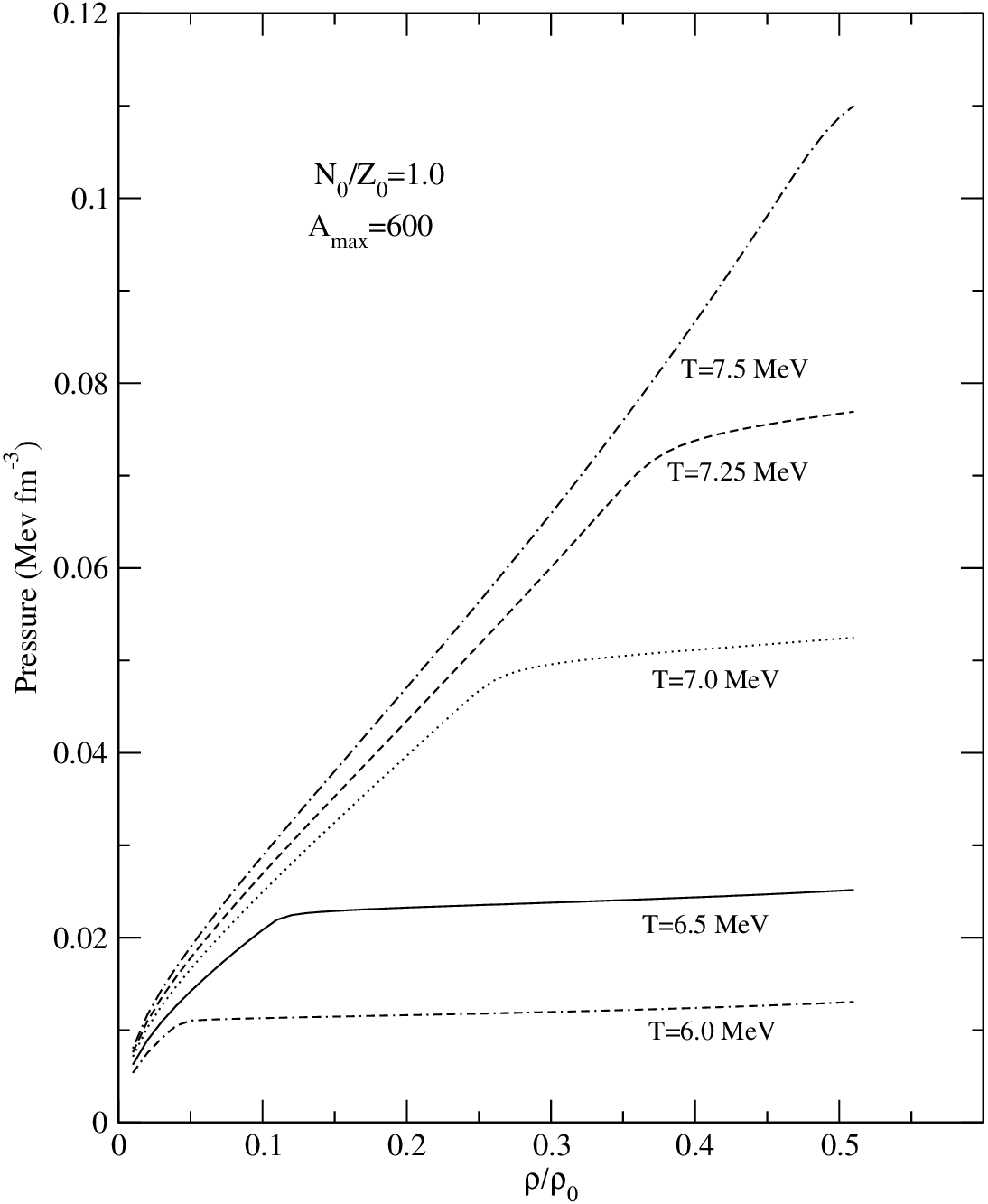}
\caption{Pressure-density isotherms at $T$=6, 6.5 ,7.0, 7.25 and 7.5 MeV for 
$N_0/Z_0=1.4$ and $A_{max}=600$. Note that the point of the beginning of
co-existence moves up and to the right as the temperature increases.}
\label{Fig. 5}
\end{figure}

\begin{figure}
\includegraphics[width=5.0in,height=6.0in,clip]{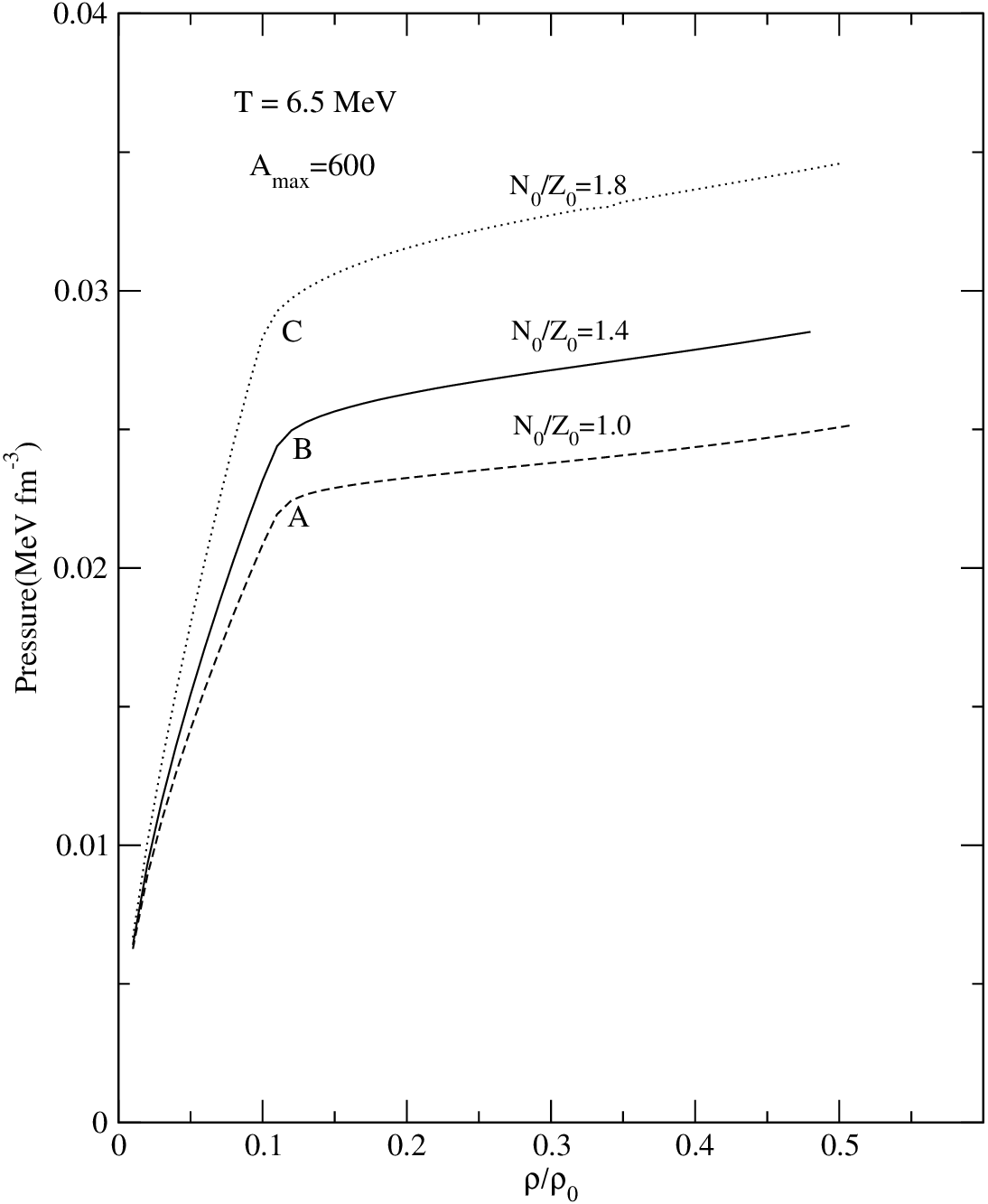}
\caption{Pressure-density curves at  $T=6.5$ MeV for three systems with 
$(N_0/Z_0)$ values equal to 1, 1.4 and 1.8. The points marked A, B and C on the
isotherms will give the values of pressure when co-exsistence sets in at 
$T=6.5$ MeV for these $N_0/Z_0$ values. }
\label{Fig. 6}
\end{figure}

\begin{figure}
\includegraphics[width=5.0in,height=5.5in,clip]{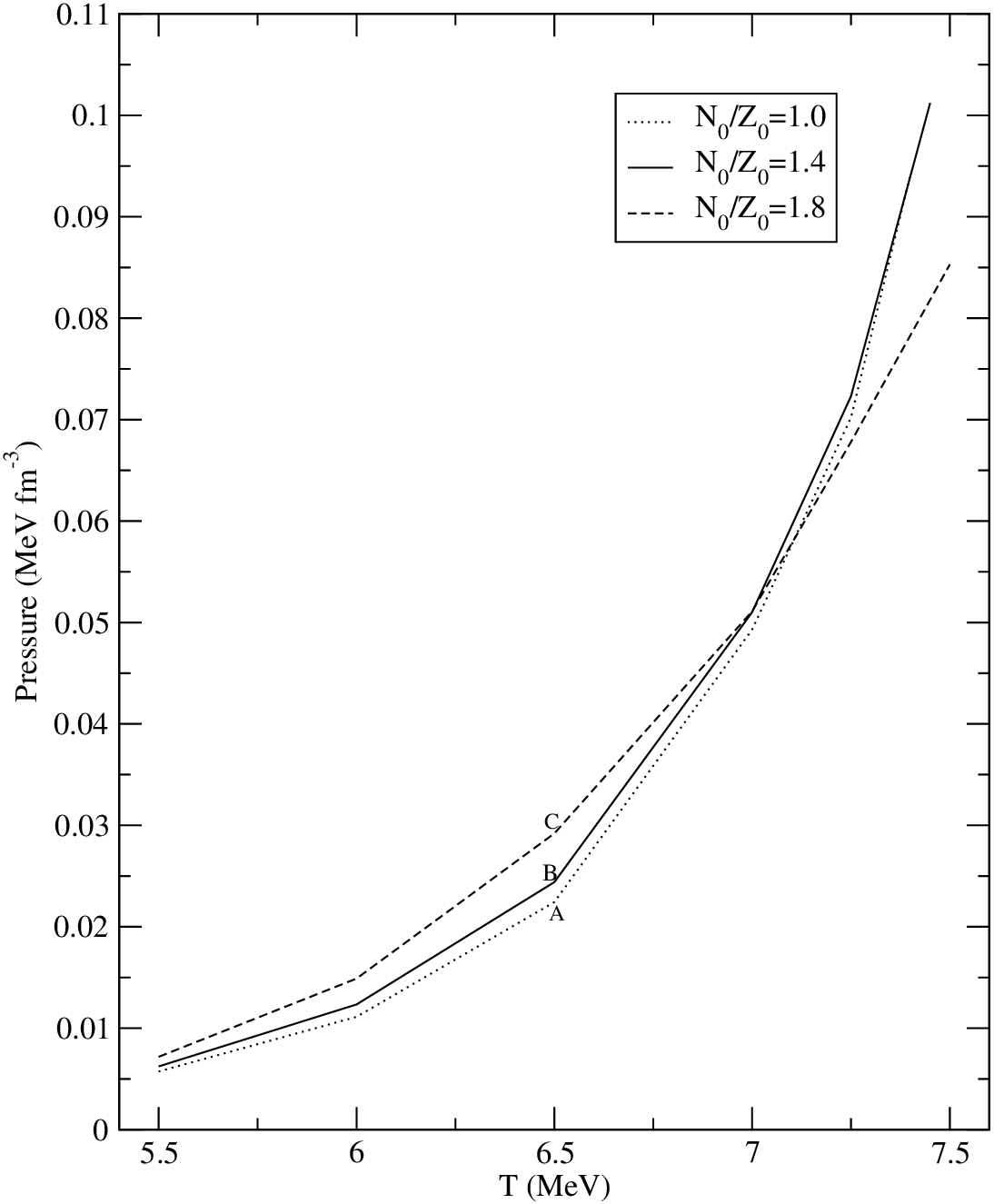}
\caption{Phase-coexistence lines in the $p-T$ plane for different values of
$N_0/Z_0$. As in Fig. 6 the points marked A, B and C gives the value
of pressure where co-existence sets in at $T$=6.5 MeV. }
\label{Fig. 7}
\end{figure}

\begin{figure}
\includegraphics[width=5.0in,height=6.0in,clip]{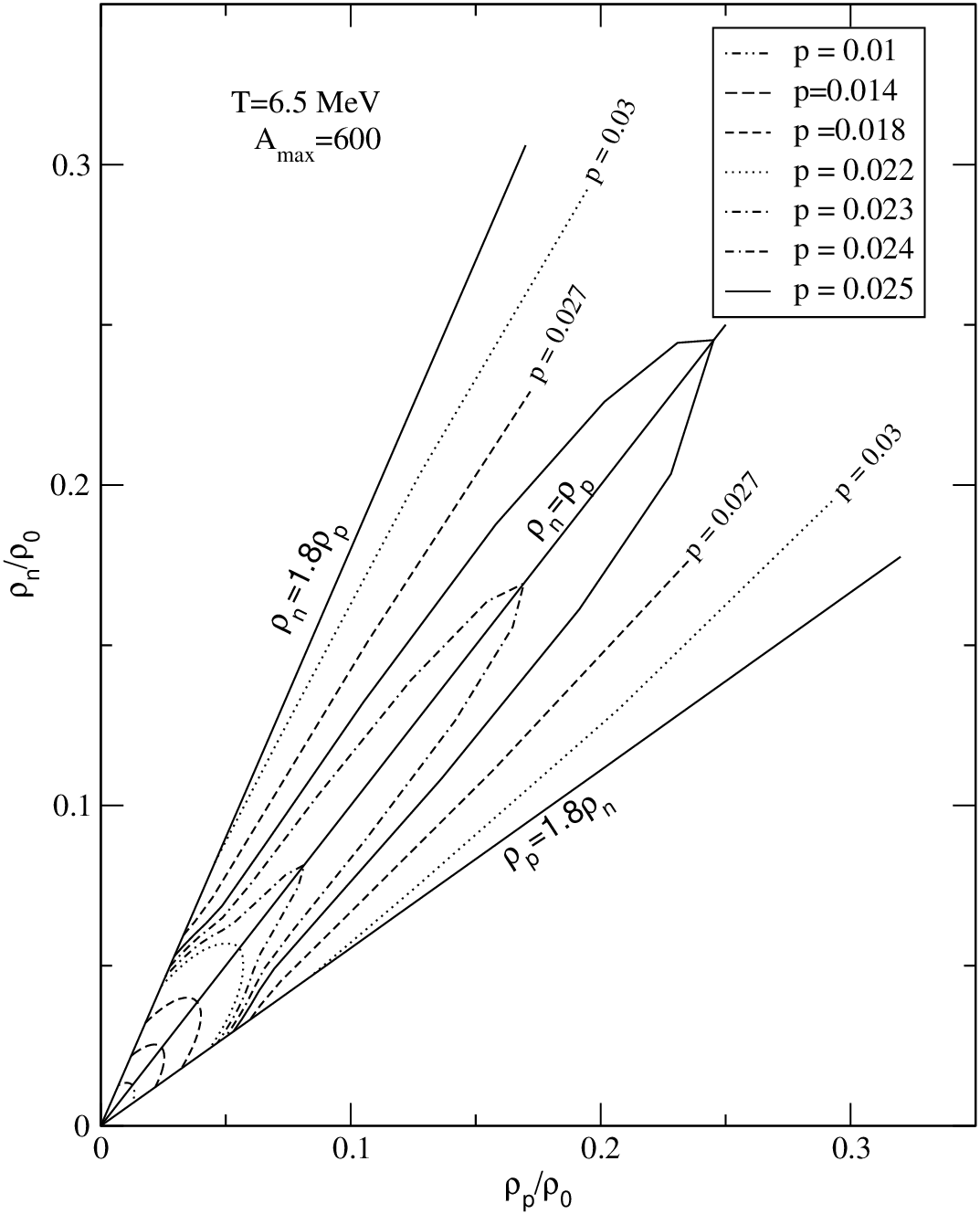}
\caption{Contours of constant pressure $p$ in $\rho_n, \rho_p$ plane at 
$T=6.5$ MeV
for $A_{max}=600$. The values of the pressure (in MeV/fm$^3$) are marked 
against the contours and
some are given in the box in the upper right corner. The region is bounded by
$\rho_n=1.8\rho_p$, $\rho_p=1.8\rho_n$ and $(\rho_n+\rho_p)/\rho_0 \leq 0.5$. 
The line $\rho_n=\rho_p$ is shown in the middle. }
\label{Fig. 8}
\end{figure}

\begin{figure}
\includegraphics[width=5.0in,height=6.0in,clip]{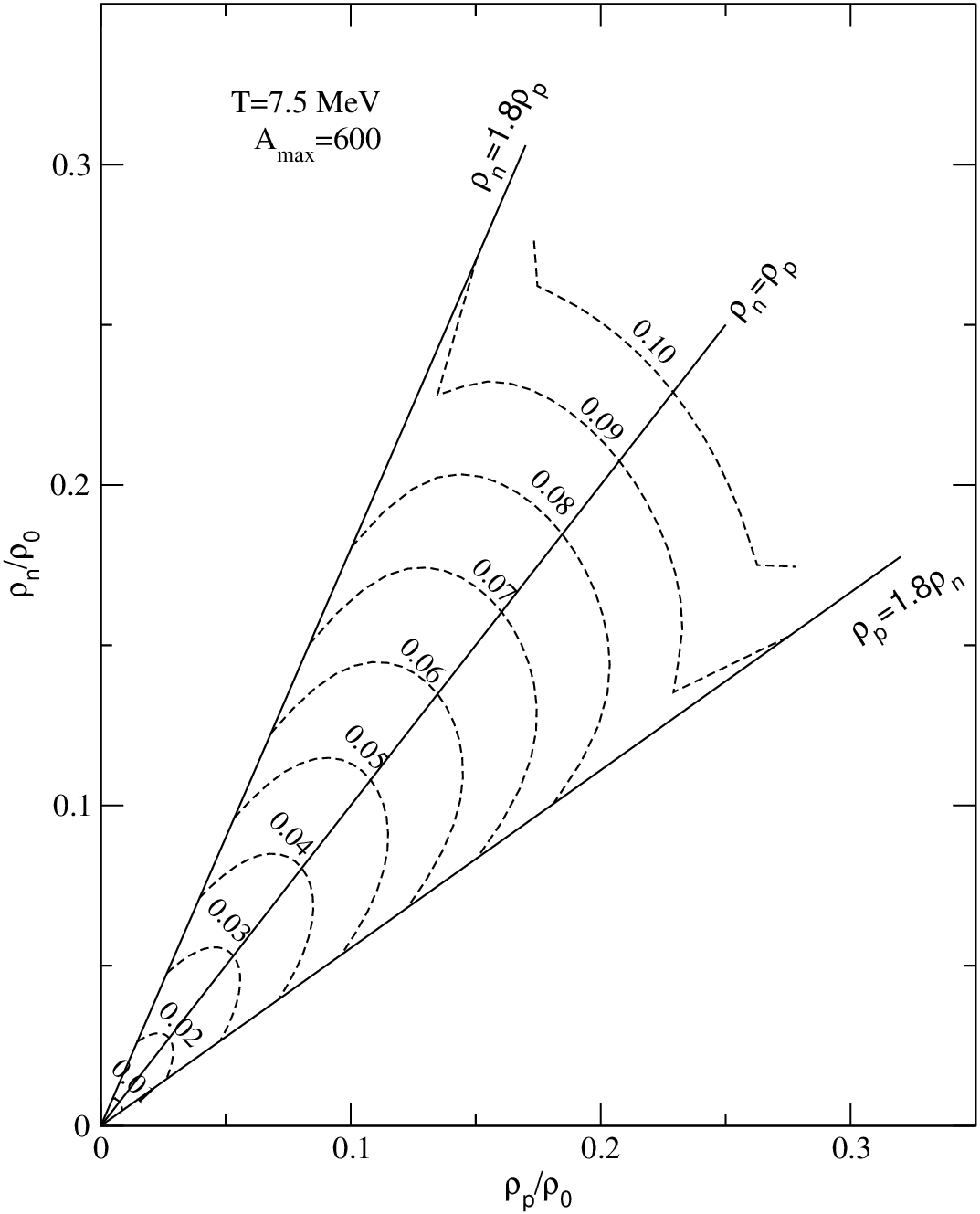}
\caption{Same as in Fig. 8 except that the temperature is 7.5 MeV. The system is
mostly in the gaseous phase which changes the shape of the contours here as
compared to that in Fig. 8. 
}
\label{Fig. 9}
\end{figure}

\begin{figure}
\includegraphics[width=5.0in,height=6.0in,clip]{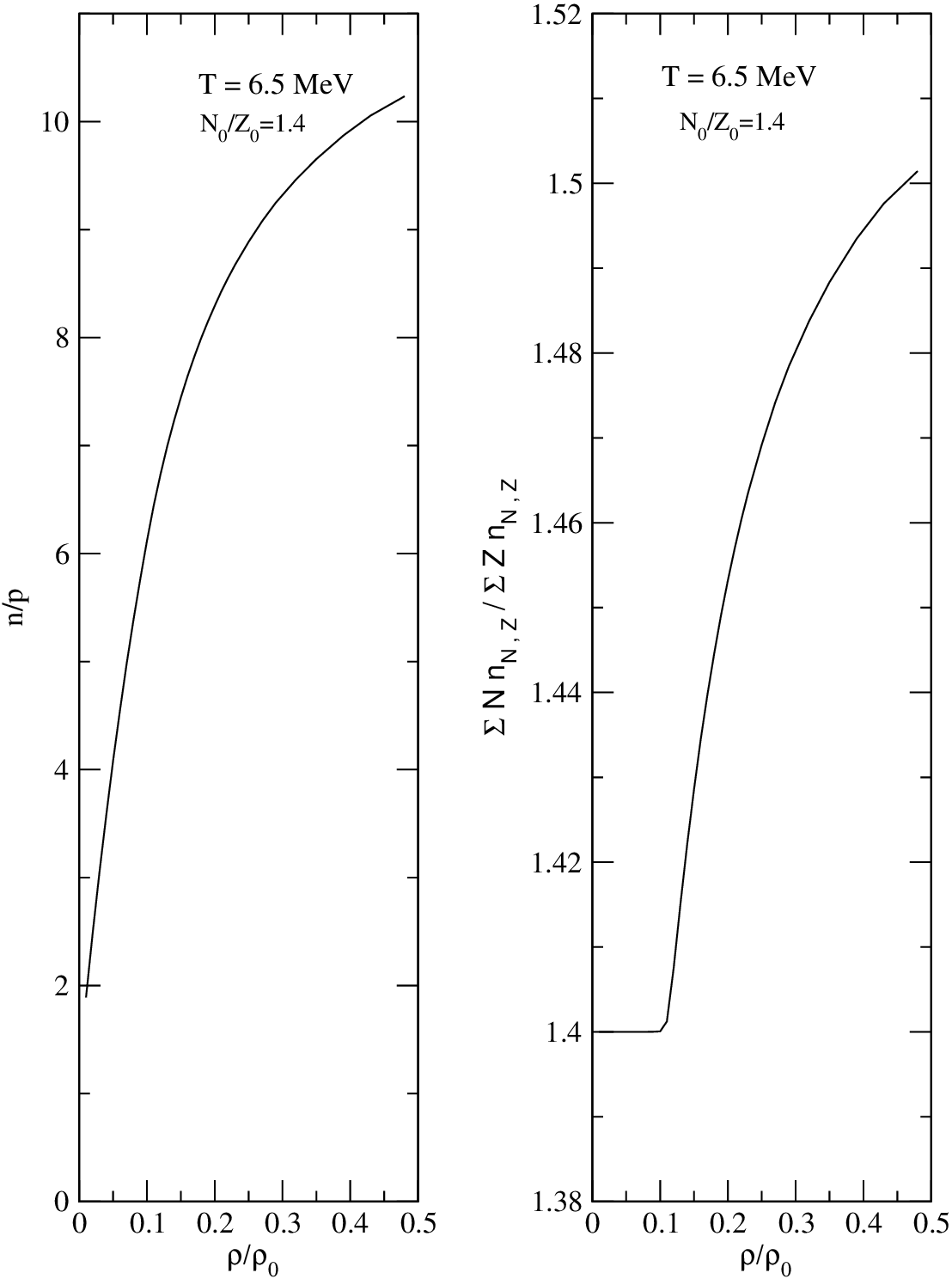}
\caption{This figure is for $T$=6.5 MeV and $N_0/Z_0=1.4$.
The left panel shows the rise of  the ratio of 
the number of free 
neutrons to the number of free protons as a function of density. 
While the rise is fast, nothing particularly new
happens at the onset of co-existence.  If, however, the gas
phase is defined to be all particles with $A\leq 70$ (this would be consistent
with Figs 2 and 3), the ratio of neutrons to protons bound in the gas phase
remains that of the parent system till co-existence sets in (right panel)
and then begins to rise.  It behaves like an order parameter if the
parent system is asymmetric. 
}
\label{Fig. 10}
\end{figure}

\end{document}